\documentclass[twocolumn]{revtex4-1}
\usepackage[dvipdfmx]{graphicx}
\usepackage{color}

\begin{document}

\title{Charge Kondo Effect and Superconductivity in the Falikov-Kimball model with the Pair Hopping}
\author{Ryu Shinzaki}
\affiliation{Department of Physics, Tokyo Institute of Technology,
Meguro, Tokyo 152-8551, Japan}
\author{Joji Nasu}
\affiliation{Department of Physics, Tokyo Institute of Technology,
Meguro, Tokyo 152-8551, Japan}
\author{Akihisa Koga}
\affiliation{Department of Physics, Tokyo Institute of Technology,
Meguro, Tokyo 152-8551, Japan}

\date{\today}

\begin{abstract}
We study the Falikov-Kimball model with the pair hopping
between the conduction and localized bands
to discuss how the charge Kondo effect is realized.
By combining dynamical mean-field theory with
the continuous time quantum Monte Carlo method,
we clarify that the charge Kondo state survives even at zero temperature
and this competes with the charge ordered and $s$-wave superconducting states.
The role of the interorbital repulsion for the superconducting state 
is also addressed.
\end{abstract}

\pacs{}


\maketitle
\section{Introduction}

Electron valence in the transition-metal and rare-earth ions has attracted
interest in the strongly correlated electron systems.
Typical examples are the valence skipping phenomena
for bismuth and thallium ions in some compounds.
In the ions,
electron configurations prefer the closed shell structure
in the $s$ orbital
 and avoid the ionic state with a spin.
This should lead to interesting low temperature properties
such as 
colossal negative thermal expansion
in La-doped BiNiO$_3$~\cite{Bi-NTE1,Bi-NTE2,Bi-NTE3} and
superconductivity in K-doped BaBiO$_3$~\cite{Bi-sys1,Bi-sys2}.
Moreover, in PbTe system with non-magnetic Tl impurities,
Kondo-like behavior appears in the resistivity,
which is known as the charge Kondo effect~\cite{Tl-ex1,Tl-ex2}. 
The valence skipping phenomenon
in $d$-electron systems has also been suggested~\cite{d-ele-sys1,d-ele-sys2},
which stimulates further theoretical investigations on the valence skipping and
related phenomena~\cite{ATT1,Bi-sys3,Taraphder,Andergassen,Tl-sys2,Naka,Kojima}.

In valence skipping ions, the effective degrees of freedom
should be represented by the empty and doubly occupied states
for the $s$ orbital.
There are two distinct models to describe the valence skip ions.
(i) in most theoretical studies,
an effective attractive interaction is introduced
in the orbital of the ions to mimic 
the stability of closed shell configurations~\cite{Bi-sys3,Tl-sys2,Taraphder,Andergassen,ATT1}.
Low temperature properties have been discussed such as
the valence transition in La-doped BiNiO$_3$~\cite{Naka,Kojima},
charge ordering and superconductivity in K-doped BaBiO$_3$~\cite{Bi-sys3},
and charge Kondo effect in Tl-doped PbTe~\cite{Tl-sys2,Andergassen}.
(ii) Another mechanism has recently been proposed,
where the interband correlations are taken into account~\cite{Matsuura}.
It has been suggested that the charge Kondo effect
in the single impurity model is well reproduced by 
the introduction of the pair hopping between
the impurity and conduction bands
in addition to the repulsive interaction.
On the other hand, as for the periodic system, 
the ground state remains unclear
as well as the finite-temperature properties.
In particular,
it should be instructive to clarify in the periodic system
the possibility of the superconductivity against the charge Kondo state
as the pair hopping may induce the superconducting (SC) 
 state, which is trivially realized in the system
with the attractive interaction~\cite{attractivePAM}.

In this paper, we study the correlated electron system
with conduction and localized bands.
By considering Coulomb interaction and pair hopping
between conduction and localized orbitals,
we discuss how the valence skipping phenomena affect low temperature properties
in the bulk system.
Here, we use dynamical mean-field theory (DMFT)~\cite{DMFT1,DMFT2,DMFT3}
combined with the continuous-time
quantum Monte Carlo (CTQMC) method~\cite{werner_hyb,ctqmc_rev}.
Examining electron configurations, charge correlations, and order parameters,
we discuss 
the stability of the charge Kondo state 
 against spontaneously
symmetry breaking states.

The paper is organized as follows.
In Sec.~\ref{MM}, we introduce the model Hamiltonian and
briefly summarize our numerical method.
In Sec.~\ref{results},
calculating various physical quantities,
we discuss the role of interorbital repulsion and pair hopping
in realizing the charge Kondo, charge ordered, and superconducting states.
Then, we determine the phase diagram.
A summary is given in the final section.

\section{Model and Method}\label{MM}
We study low energy properties in strongly correlated electron systems
with the localized valence skipping ions.
To this end, we deal with the extended Falikov-Kimball model~\cite{FK,FK_rev},
where conduction electrons interact with localized ones.
This is the natural extension of the impurity model
discussed in Ref.~\cite{Matsuura}, and its Hamiltonian is given as,
\begin{eqnarray}
H&=&H_0+H',\\
H_0&=&\sum_{ij\sigma}\left(
t_{ij}c_{i\sigma}^{\dagger}c_{j\sigma}
+\epsilon_d\delta_{ij}n^d_{i\sigma}\right),\\
H'&=&U_{cd}\sum_{i\sigma\sigma'}n^c_{i\sigma}n^d_{i\sigma'}\nonumber\\
&&-J_{ph} \sum_i \left( c_{i\downarrow}^\dag c_{i\uparrow}^\dag
d_{i\uparrow} d_{i\downarrow} +H.c. \right),\label{eq:Hami}
\end{eqnarray}
where $c_{i\sigma}$ ($d_{i\sigma}$) is an annihilation operator
of a conduction electron (localized electron)
with spin $\sigma(=\uparrow,\downarrow)$.
$n^c_{i\sigma}(=c^\dagger_{i\sigma}c_{i\sigma})$ and
$n^d_{i\sigma}(=d^\dagger_{i\sigma}d_{i\sigma})$ are
the number operators of the conduction
and localized electrons at the $i$th site, respectively.
$t_{ij}(=-t\delta_{\langle ij\rangle})$ is the hopping integral
of the conduction electrons between the nearest-neighbor sites
and $\epsilon_d$ is the energy level of the $d$ orbitals.
$U_{cd} (J_{ph})$ is the repulsive interaction (pair hopping)
between the conduction and localized electrons.

When $J_{ph}=0$, the system is reduced to
the conventional Falikov-Kimball model~\cite{FK}.
In the infinite dimensions~\cite{FK_rev}, the model is exactly solved, and
ground state properties have been discussed in detail~\cite{chung_2000}.
It is known that, in the presence of the particle-hole symmetry,
the interorbital Coulomb interaction suppresses the single occupancy 
at each orbital
and the charge ordered (CO) state is realized at zero temperature.

In the paper, we consider both the interorbital Coulomb interaction and
pair hopping between the conduction and localized bands
on an equal footing.
In the atomic limit $(t_{ij}=0)$ under the particle-hole symmetry,
these interactions prefer the electronic configuration
with one of two orbitals empty and the other doubly occupied, while
no singly occupied states are realized in each orbital.
Therefore, valence skip feature should be captured in our model.
To provide more insight,
we wish to introduce the pseuedo spin operators for $\alpha$th band as,
\begin{eqnarray}
I_{i\alpha}^x&=&\frac{1}{2}\left(\alpha_{i\uparrow}^\dag \alpha_{i\downarrow}^\dag+\alpha_{i\downarrow} \alpha_{i\uparrow}\right),\\
I_{i\alpha}^y&=&\frac{1}{2i}\left(\alpha_{i\uparrow}^\dag \alpha_{i\downarrow}^\dag-\alpha_{i\downarrow} \alpha_{i\uparrow}\right),\\
I_{i\alpha}^z&=&\frac{1}{2}(n_i^\alpha-1).
\end{eqnarray}
Then, the interaction part of the original Hamiltonian can be rewritten
as the following the Kondo lattice model 
with anisotropic interactions~\cite{Kikuchi}:
\begin{eqnarray}
H'&=&2\sum_i \Big[2U_{cd}I_{ic}^zI_{id}^z
+J_{ph}\left(I_{ic}^xI_{id}^x+I_{ic}^yI_{id}^y\right)\Big].\label{II}
\end{eqnarray}
We wish to note that $U_{cd}$ and $J_{ph}$ yield distinct low temperature
properties.
When $|J_{ph}|\ll 2U_{cd}$, the diagonal Ising interactions
make the pseuedo-spins antiparallel in the $z$ direction.
If one considers the lattice model,
the antiferro-type ordered state is realized with
the staggered pseuedo-spin moments $\langle I_{i\alpha}^z\rangle\sim 
(-1)^{i+\delta_\alpha}$,
where $\delta_\alpha=0 (1)$ for conduction (localized) band.
This implies that the CO state is realized in the original model.
The characteristic quantities are alternating electron densities
$\rho_c$ and $\rho_d$, where
$\rho_\alpha=\sum_{i}(-1)^in_{i}^\alpha/N$.
In the opposite case with $|J_{ph}|\gg 2U_{cd}$,
the pseuedo-spins are on the $xy$ plane with the staggered configuration, 
{\it e.g.},
$\langle I_{i\alpha}^x\rangle\sim (-1)^{i+\delta_\alpha}$
due to the inplane anisotropy in Eq.~(\ref{II}).
Then the superconducting state is realized with the staggered pair potential
$\langle \alpha_{i\uparrow}\alpha_{i\downarrow}\rangle \sim (-1)^{i+\delta_\alpha}$.
When the particle-hole transformation
$c_{i\sigma}\rightarrow (-1)^i \sigma \tilde{c}_{i\sigma}^\dag$ is applied,
$H(t, U_{cd}, J_{ph})$ is transformed to $H(t, U_{cd}, -J_{ph})$,
and the superconducting order parameter is uniform in the model.
Therefore, the sign of the pair hopping is essentially irrelevant,
and the SC state can be regarded as a conventional $s$-wave SC state.
When $2U_{cd}=J_{ph}$,
the system is reduced to the isotropic Kondo lattice model.
In the strong coupling case,
the Kondo insulating state is realized with the pseuedo-spin singlet
($\langle {\bf I}_c\cdot {\bf I}_d\rangle=-3/4$).
This implies the existence of the charge Kondo state in our model,
which is mainly formed by empty and doubly occupied states.

To study the competition between the SC, CO, and charge Kondo states
in the original model Eq.~(\ref{eq:Hami}),
we make use of DMFT~\cite{DMFT1,DMFT2,DMFT3}
in the Nambu formalism~\cite{NambuDMFT}.
In the framework of DMFT, the lattice model is mapped to an effective
impurity model, where local electron correlations
are taken into account precisely.
The Green function for the original lattice system is then obtained
via self-consistency equations imposed on the impurity problem.
The non-interacting Green function in the lattice system is represented as 
the two-by-two matrix,
\begin{eqnarray}
\hat{G}_{0\alpha}({\bf k},i\omega_n)=
\Big[i\omega_n\hat{\sigma}_0+(\mu-\epsilon_{\alpha {\bf k}})\hat{\sigma}_z\Big]^{-1},
\end{eqnarray}
where $\hat{\sigma}_0$ is the identity matrix, $\hat{\sigma}_z$ is the $z$
component of the Pauli matrix, $\omega_n=(2n+1)\pi T$ with interger $n$ 
is the Matsubara
frequency, $T$ is the temperature, and $\mu$ is the chemical potential.
$\epsilon_{\alpha {\bf k}}$ is the dispersion relation for
the $\alpha(=c, d)$th band, namely,
$\epsilon_{c{\bf k}}=\epsilon_{\bf k}$ and $\epsilon_{d{\bf k}}=\epsilon_d$.
Since there is no hybridization between conduction and localized bands, 
no interband elements appear in the Green's function~\cite{OSMT}.
The lattice Green's function
is then given by the site-diagonal selfenergy as
\begin{equation}
\hat{G}_\alpha(i\omega_n)=\int d{\bf k}
\Big[\hat{G}_{0\alpha}^{-1}({\bf k},i\omega_n)
-\hat{\Sigma}_\alpha(i\omega_n)\Big]^{-1},
\end{equation}
where the Green's functions and selfenergy are
represented in the Nambu formalism.

In the following, we use the semicircular density of states
$\rho(x)=2\sqrt{1-(x/D)^2}/\pi D$,
which corresponds to an infinite-coordinate Bethe lattice.
By using Dyson equations, the self-consistency condition
is represented by the Green's function of the conduction bands, as
\begin{eqnarray}
\hat{\cal G}^{-1}(i\omega_n)=i\omega_n\hat{\sigma}_0+\mu\hat{\sigma}_z
-\left(\frac{D}{2}\right)^2\hat{\sigma}_z \hat{G}_c(i\omega_n)\hat{\sigma}_z,
\label{eq:self}
\end{eqnarray}
where $\hat{\cal G}$ is the non-interacting Green function
in the effective impurity model.

There are various numerical methods to solve the effective impurity problem.
To discuss quantitatively
how the SC and CO states compete with the charge Kondo state,
we use here the CTQMC method~\cite{werner_hyb,ctqmc_rev}.
In our model, the double expansion technique~\cite{Steiner},
where the partition function is expanded
with respect to both the effective bath and the pair hopping,
is efficient to perform Monte Carlo simulations without minus sign problems.
In the paper, to discuss how the valence skipping phenomenon is realized,
we evaluate the probabilities of
empty, singly, and doubly occupied states in each orbital as
$
\langle e_{i\alpha} \rangle, \langle s_{i\alpha\sigma} \rangle$,
and $\langle d_{i\alpha} \rangle
$,
where $e_{i\alpha}=(1-n_{i\uparrow}^\alpha)(1-n_{i\downarrow}^\alpha)$,
$s_{i\alpha\sigma}=n_{i\sigma}^\alpha(1-n_{i\bar{\sigma}}^\alpha)$, and
$d_{i\alpha}=n_{i\uparrow}^\alpha n_{i\downarrow}^\alpha$, respectively.
In the following, we take $D$ as unit of energy
and set $\mu=U_{cd}$ and $E_d=0$ to discuss low temperature properties
in the system with particle-hole symmetric conditions 
$\langle n_c\rangle=\langle n_d\rangle=1$.

\section{results}\label{results}

\begin{figure}[htb]
\begin{center}
\includegraphics[width=8cm]{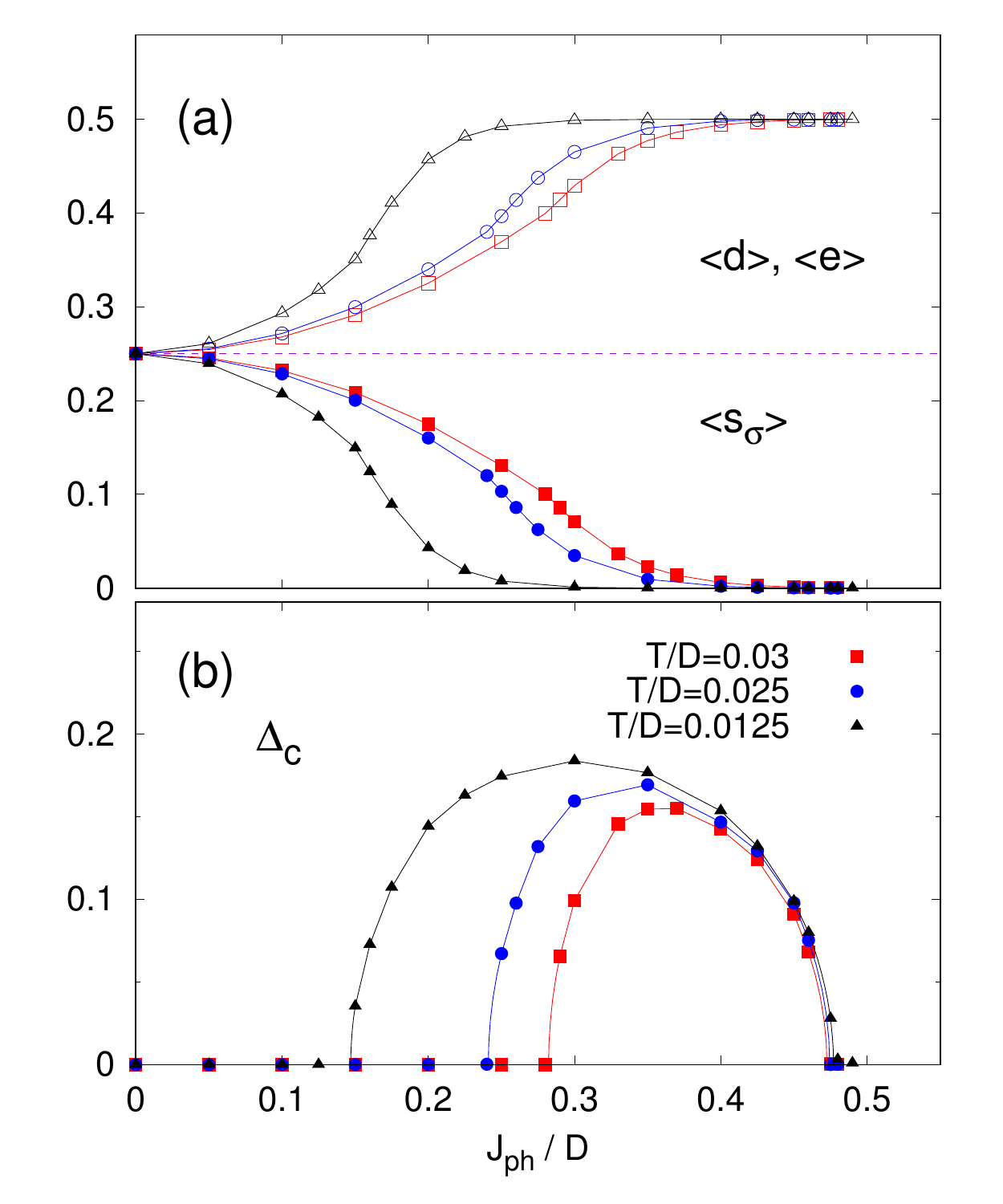}
\end{center}
\caption{
(a) electron configurations
$\langle d \rangle, \langle s_\sigma \rangle$, and $\langle e \rangle$,
and (b) the pair potential in the system with $U_{cd}=0$ at the temperature
$T/D=0.03, 0.025$ and $0.0125$, respectively.
Dashed lines are guides to the eye.
}
\label{fig:Delta-U0}
\end{figure}

\begin{figure}[htb]
\begin{center}
\includegraphics[width=8cm]{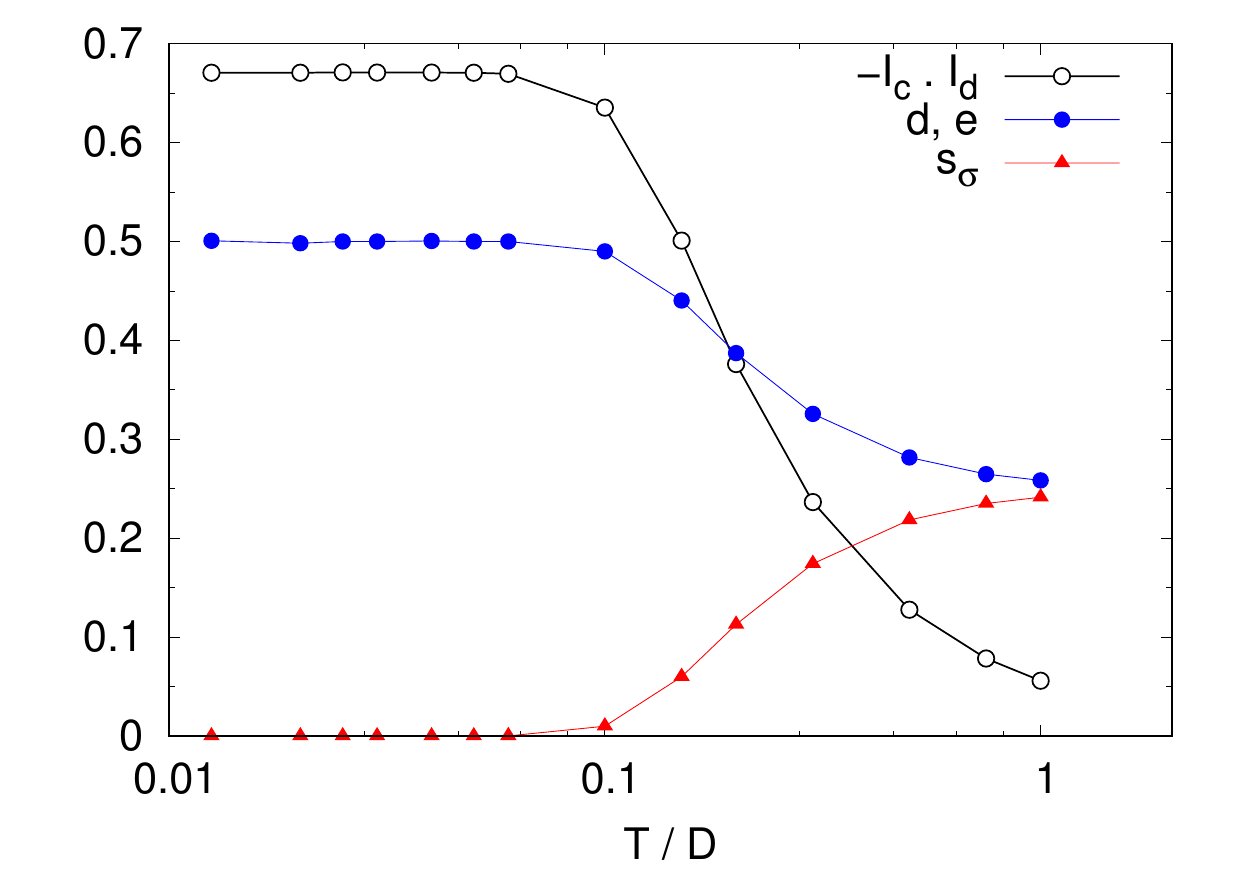}
\end{center}
\caption{
Pseuedo spin correlations $\langle{\bf I}_c\cdot {\bf I}_d\rangle$ and
probabilities as a function of the temperature
in the system with $U_{cd}=0$ and $J_{ph}/D=0.75$.
}
\label{fig:IcId}
\end{figure}

We discuss low temperature properties in the system
with itinerant and localized bands.
Fixing the interorbital Coulomb interaction as $U_{cd}=0$,
we focus on the effect of the pair hopping in the system
to discuss the competition between the charge Kondo and SC states.
We first calculate the probabilities of empty, singly,
and doubly occupied states
to examine the electron configuration in the system.
The results are shown in Fig.~\ref{fig:Delta-U0}(a).
When $J_{ph}=0$, the system is noninteracting, and the metallic state is
realized with
$\langle e^c \rangle =\langle s^c_{\sigma} \rangle =
\langle d^c\rangle =0.25$.
The introduction of the pair hopping
increases the probabilities of empty and double occupied states, 
while it decreases those of single occupied states.
In the strong coupling limit,
these values $\langle e^c\rangle =\langle d^c \rangle\rightarrow 0.5$
and $\langle s^c_\sigma\rangle\rightarrow 0$.
Similar behavior appears in the localized bands (not shown), 
which means that the valence skip behavior is well described
by the pair hopping.
To discuss how the SC state is realized in the system,
we also calculate the pair potential in the conduction band
$\Delta_c=\langle c_\uparrow c_\downarrow\rangle$,
as shown in Fig.~\ref{fig:Delta-U0}(b).
In the intermediate coupling region $(J_{ph})_{c1}<J_{ph}<(J_{ph})_{c2}$,
the SC state is realized with a finite pair potential.
Namely, the phase transitions are of second order and
the critical interactions are deduced as
$(J_{ph})_{c1}/D\sim 0.28$ and
$(J_{ph})_{c2}/D\sim 0.47$ at $T/D=0.03$.
An important point is that
electron configurations are gradually changed in the SC state,
as shown in Fig.~\ref{fig:Delta-U0}(a).
Around $(J_{ph})_{c1}$, 
the single occupancy still appears because of weak electron correlations. 
In the case, the BCS-like SC state is realized, and thereby 
the critical value $(J_{ph})_{c1}$ approaches zero 
with decreasing temperatures,
as shown in Fig.~\ref{fig:Delta-U0}(b).
Roughly speaking, 
we find that the emergence of the phase transition appears 
to be related with the double occupancy of $\langle d \rangle \sim 0.4$, 
which helps us to discuss later the effect of the interorbital interaction.
On the other hand, in the stronger coupling region
singly occupied states are little realized,
as shown in Fig.~\ref{fig:Delta-U0}(a).
Therefore, paired electrons formed by pair hopping play a crucial role
in the region.
When paired electrons are itinerant in the lattice $[J_{ph}<(J_{ph})_{c2}]$,
the SC state is realized.
On the other hand, when $J_{ph}>(J_{ph})_{c2}$,
the paired electrons are localized in each site,
which is expected to correspond to the charge Kondo state.
To clarify whether or not the charge Kondo state is realized
at low temperatures,
we show in Fig.~\ref{fig:IcId}
the temperature dependence of the pseuedo spin correlation
$\langle {\bf I}_c\cdot {\bf I}_d\rangle$
for the system with $U_{cd}=0$ and $J_{ph}/D=0.75$.
Its magnitude becomes larger
with decreasing temperature at $T\sim |J_{ph}|$.
The quantity is almost saturated below $T/D\sim 0.1$,
where $\langle e^c\rangle = \langle d^c\rangle \sim 0.5$ and
$\langle s^c_\sigma\rangle \sim 0$.
In addition, we could not find the SC state at lower temperatures,
suggesting that
the charge Kondo state is realized even at zero temperature.
This is consistent with
the fact that the critical point $(J_{ph})_{c2}$ 
between the SC and charge Kondo states
is little changed with decreasing temperature,
as shown in Fig.~\ref{fig:Delta-U0}(b).

\begin{figure}[htb]
\begin{center}
\includegraphics[width=8cm]{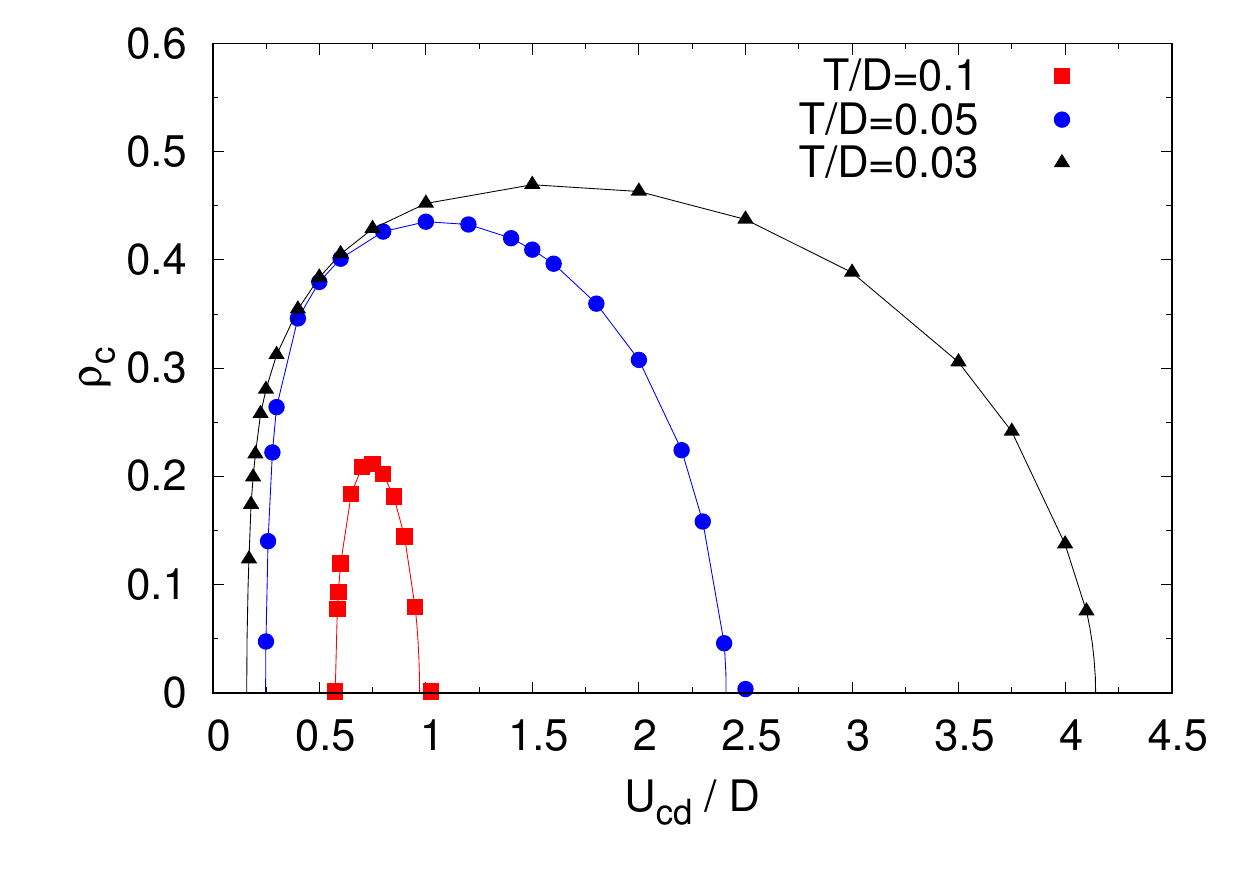}
\end{center}
\caption{
Solid squares, circles, and triangles represent order parameters
for the CO state in the system with $J_{ph}=0$ at the temperature
$T/D=0.1, 0.05$, and $0.03$, respectively.
}
\label{fig:rho-J0}
\end{figure}

We next consider how the interorbital interaction $U_{cd}$ stabilizes
the CO state~\cite{CO1,CO2,CO3,CO4}.
When $J_{ph}=0$, the system is reduced to a conventional Falikov-Kimball
model and its low temperature properties have been discussed 
in detail~\cite{FK,FK_rev}.
Figure~\ref{fig:rho-J0} shows the order parameter at fixed temperatures
$T/D=0.03, 0.05$ and $0.1$.
We find that the CO state is realized in the intermediate coupling region
$U_{cd}/D\sim 1$
and becomes more stable with decreasing temperatures.
This is consistent with the fact that the CO state is always a ground state
in the system without the pair hopping $J_{ph}=0$~\cite{CO1,CO2,CO3,CO4}.

\begin{figure}[htb]
\begin{center}
\includegraphics[width=8cm]{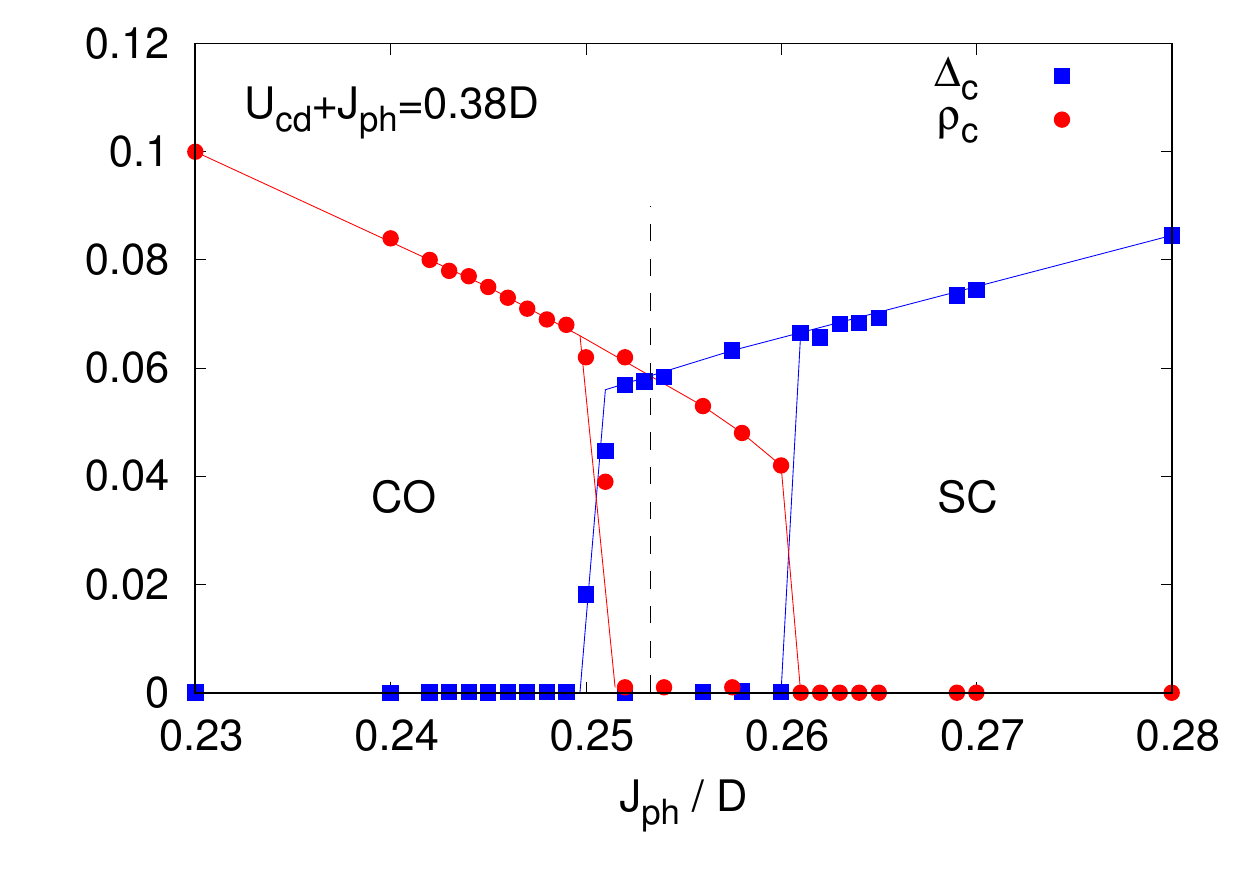}
\end{center}
\caption{
Circles and squares represent order parameters for the SC and CO states
when $U_{cd}+J_{ph}=0.38D$ and $T/D=0.03$.
Solid lines serve as a guide to the eye and
dashed line represents the symmetric point ($J_{ph}/D=0.2533$).
}
\label{fig:line38}
\end{figure}

From these results in two limiting cases,
we find two distinct ordered states.
Now, another question arises how the SC and CO states
compete with each other.
Here, we fix the condition $U_{cd}+J_{ph}=0.38D$ to clarify
how these two phases are realized.
Figure~\ref{fig:line38} shows order parameters in the system at $T/D=0.03$.
When $J_{ph}/D=0.23$, the CO state is realized with the order parameter
$\rho_c\sim 0.1$.
The increase of the pair hopping monotonically decreases this quantity up to
$J_{ph}/D\sim 0.26$.
The order parameter suddenly vanishes and
the finite pair potential appears instead.
This implies the existence of the first-order phase transition between
CO and SC states.
In fact, the SC state solution exists in the case $J_{ph}/D\gtrsim 0.25$
shown as the solid squares in Fig.~\ref{fig:line38}.
Note that
at the symmetric point ($2U_{cd}=J_{ph}=0.2533D$),
order parameters take the same value within our numerical accuracy.
This originates from the fact that
the Hamiltonian Eq.~(\ref{II}) is isotropic and
these two states are degenerate at zero temperature.
Then, we conclude that there exists the first-order phase boundary along
the symmetric condition at low temperatures.

\begin{figure}[htb]
\begin{center}
\includegraphics[width=8cm]{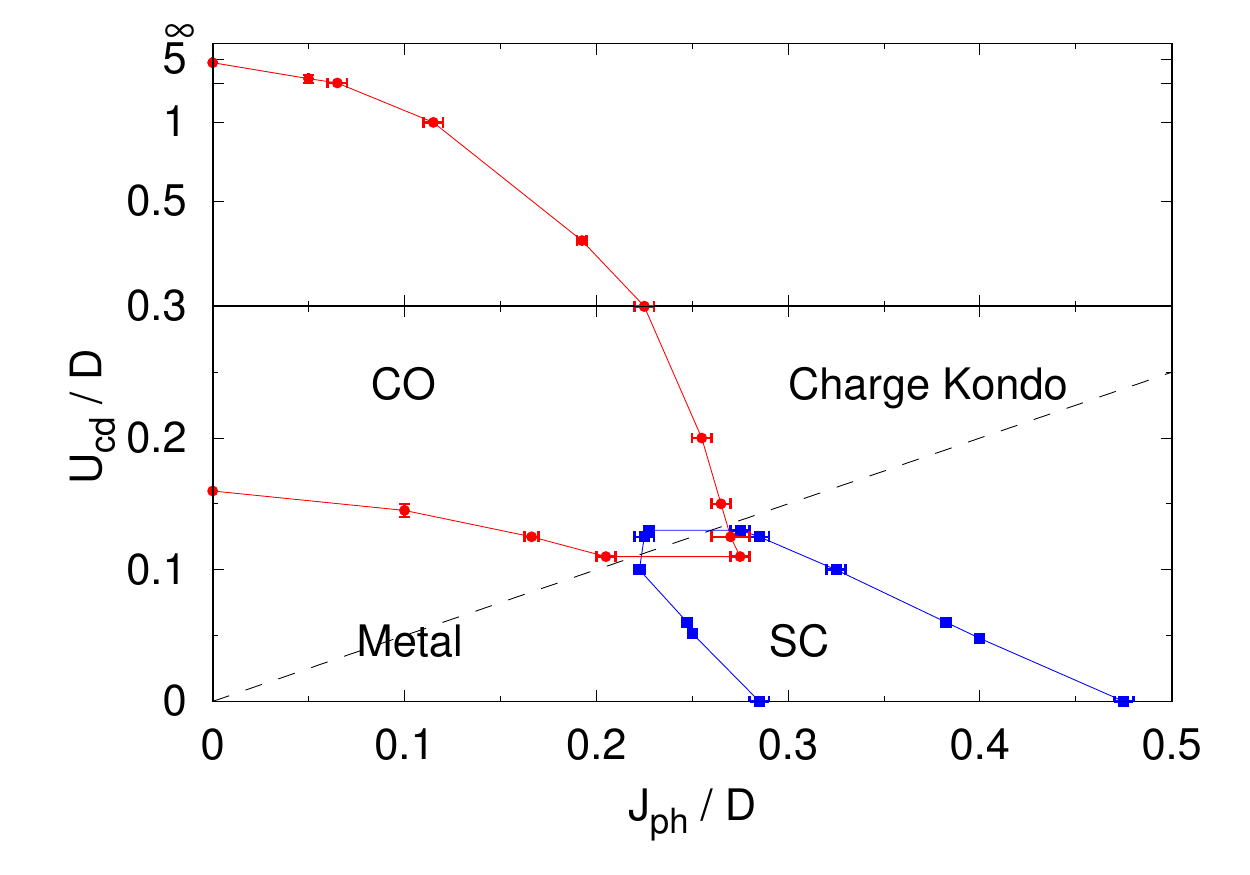}
\end{center}
\caption{
Phase diagram of the system when $T/D=0.03$.
}
\label{fig:PD3}
\end{figure}

\begin{figure}[htb]
\begin{center}
\includegraphics[width=8cm]{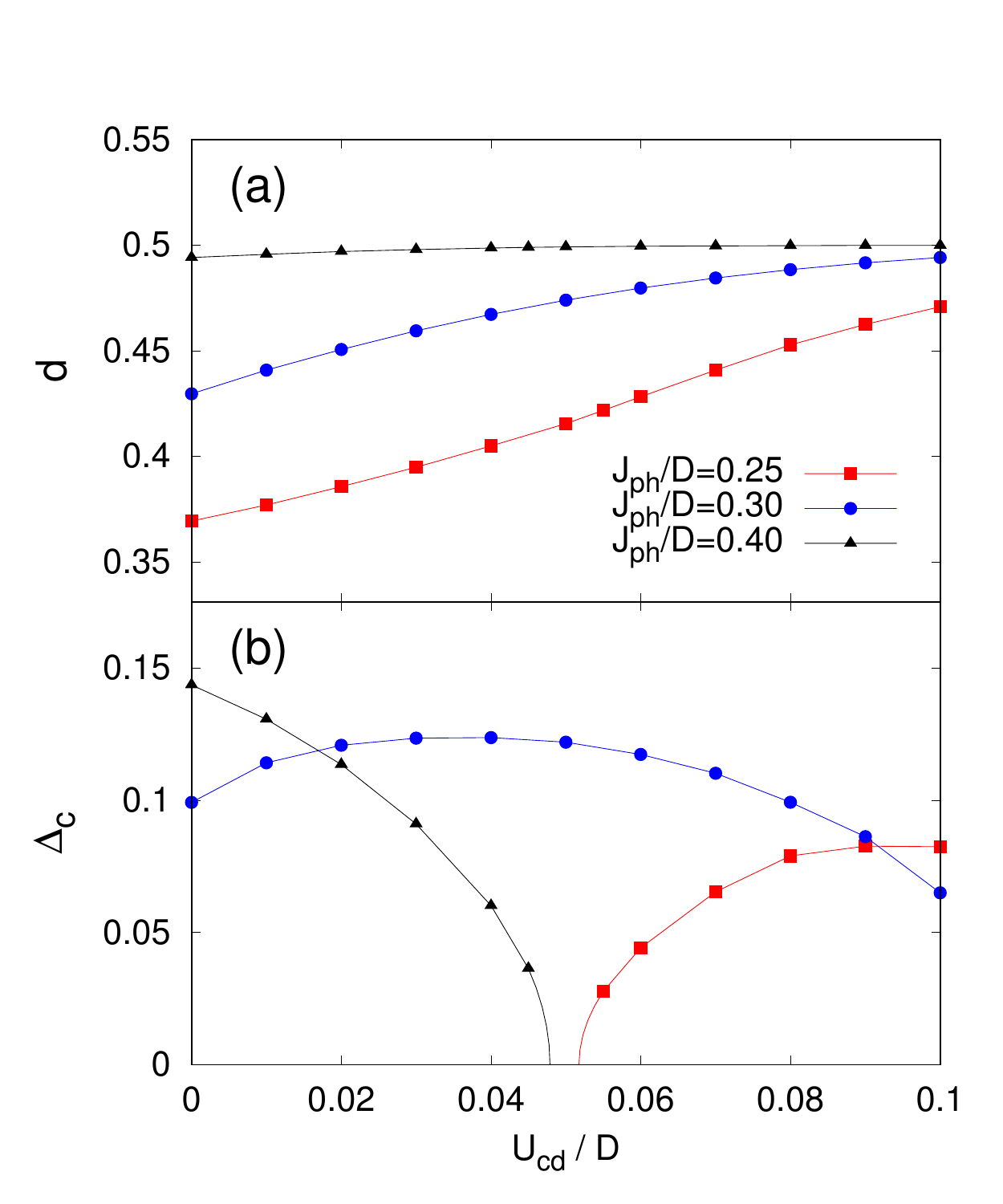}
\end{center}
\caption{
(a) Double occupancy 
$\langle d\rangle (=\langle e\rangle = (1-\langle s_\sigma \rangle)$
and (b) pair potential $\Delta_c$ as a function of $U_{cd}/D$.
Solid squares, circles, and triangles represent the results for
the systems with $J_{ph}/D=0.25, 0.3$ and $0.4$ at $T/D=0.03$.
}
\label{fig:newfig}
\end{figure}

By performing similar calculations,
we obtain the phase diagram with a fixed temperature $T/D=0.03$,
as shown in Fig.~\ref{fig:PD3}.
When the system is weakly correlated with $U_{cd}, J_{ph}\ll D$,
the metallic state is realized.
The pair hopping term induces the $s$-wave SC state,
while the interorbital interaction induces the CO state.
These two solutions overlap
around the symmetric line, as discussed above.
We also find that the phase boundary for the metallic and SC states
becomes lower when the interorbital interaction is switched on.
This should be explained by the fact that, 
in the weak coupling region, the SC state is induced when empty and doubly
occupied states become dominant.
In fact, when $J_{ph}/D=0.25$ and $U_{cd}=0$,
the introduction of the interorbital Coulomb interaction
increases the double occupancy while suppresses single occupancy,
as shown in Fig.~\ref{fig:newfig}(a).
We find that the SC state is induced around 
$U_{cd}/D\sim 0.052$ $(\langle d\rangle\sim 0.4)$.
We wish to note that the instability can not be described 
in the BCS theory, where the interorbital Coulomb interaction 
has little effects on the SC state.
Therefore, we can say that dynamical correlations play an important role
in stabilizing the SC state.
This is similar to the SC state
in the repulsive Hubbard model with degenerate orbitals~\cite{KogaSC},
where the interband Coulomb interaction induces the SC state
in a certain region.
These discussions are also applied to the CO state in the weak coupling region,
where the pair hopping $J_{ph}$ play a role in forming paired electrons.

In the strong coupling region with $U_{cd}, J_{ph}\gg D$,
the paramagnetic state appears in the phase diagram.
Decreasing temperatures, the state is adiabatically connected to
the charge Kondo state.
In the large $U_{cd}$ case, the characteristic energy for the CO state
is $D^2/U_{cd}$, while that for charge Kondo state
is the pair hopping $J_{ph}$.
Therefore, the ground-state phase boundary 
between the CO and charge Kondo states
should be scaled as $\sim U_{cd}J_{ph}/D^2$
and infinitesimal $J_{ph}$ induces the charge Kondo state
in the $U_{cd}\rightarrow \infty$ limit.
On the other hand, in the $U_{cd}=0$ case, the charge Kondo state
competes with the SC state and its phase boundary is at finite $J_{ph}$
in the ground state, as discussed before.
Switching $U_{cd}$ enhances the pseuedo spin correlations,
stabilizing the charge Kondo state.
Therefore, the introduction of the interorbital Coulomb interaction makes
the strong-coupling SC state unstable,
which is clearly found in the case $J_{ph}/D=0.4$ in Fig.~\ref{fig:newfig}(b).
Then, the linear behavior in the phase boundary between
the SC and charge Kondo states appears in the strong coupling region.

Before closing this paper,
we comment on the effect of the single electron hopping (hybridization) between
the conduction and localized band, which has been treated
in the periodic Anderson model.
Since the self-consistency condition Eq.~(\ref{eq:self})
is not changed~\cite{PAM},
one can treat this model in the same framework
to discuss the possibility of the magnetically ordered state and 
competition between magnetic and charge Kondo states~\cite{KogaPeters}.
However, the single hopping gives rise to minus sign problems in
solving the effective impurity model by means of the CTQMC method.
Therefore, the quantitative analysis should be restricted
at relatively higher temperatures.
Furthermore, this single hopping makes singly occupied states active and
the nature of the valence skip ions becomes obscure,
which is beyond the scope of our paper.
Therefore, this interesting question is left for future work.

\section{Conclusion}\label{conclusion}
We have investigated the extended Falikov-Kimball model with
the Coulomb and pair hopping between the conduction and localized bands
to discuss how the valence skipping ions induce spontaneously
symmetry breaking state.
By combining DMFT 
 with
the CTQMC 
 method,
we have determined the finite temperature phase diagram, where
the SC and CO states compete with the charge Kondo state.
It is found that, in the weak coupling region, the Coulomb interaction assists
the stability of the SC state, which is a common feature inherent in
the multiorbital systems.


\begin{acknowledgements}
The authors would like to thank S. Hoshino and P. Werner
for valuable discussions.
Parts of the numerical calculations are performed
in the supercomputing systems in ISSP, the University of Tokyo.
The simulations were performed using some of the ALPS libraries~\cite{alps2}.
This work was partly supported by the Grant-in-Aid for
Scientific Research from JSPS, KAKENHI Grant Number
JP16K17747, JP16H02206, and JP16H00987 (J.N.), and
JP17K05536 and JP16H01066 (A.K.).

\end{acknowledgements}

\bibliography{./refs}

\end{document}